\documentclass[conference]{IEEEtran}
\IEEEoverridecommandlockouts

\usepackage{mathtools}	
\usepackage{amssymb}    
\usepackage{amsthm}

\usepackage[utf8]{inputenc}

\usepackage{ifthen}

\usepackage{microtype}

\usepackage{caption}

\usepackage{bm}

\usepackage[draft,nomargin,marginclue,inline]{fixme}
\makeatletter
\renewcommand*\FXLayoutMarginClue[3]{%
  \marginpar[%
  \raggedleft\@fxuseface{margin}\textcolor{red}{\ignorespaces $ \Rightarrow $}]{%
    \raggedright\@fxuseface{margin}\textcolor{red}{\ignorespaces $ \Leftarrow $}}}
\makeatother	

\usepackage{tumcolor}

\usepackage{pgfplotstable}	
\pgfplotsset{compat=1.15}

\pgfplotsset{
	discard if/.style 2 args={
        x filter/.append code={
            \edef\tempa{\thisrow{#1}}
            \edef\tempb{#2}
            \ifx\tempa\tempb
                \def\pgfmathresult{inf}
            \fi
        }
    },
    discard if not/.style 2 args={
        x filter/.append code={
            \edef\tempa{\thisrow{#1}}
            \edef\tempb{#2}
            \ifx\tempa\tempb
            \else
                \def\pgfmathresult{inf}
            \fi
        }
    }
}

\usepackage{cuted}

\usepackage{url}
\usepackage{cite}

\usepackage{textcomp}

\usepackage{subcaption}

\usepackage{cleveref}

\usepackage{enumitem}

\usepackage{algorithm}
\usepackage{algorithmic}

\newlength{\leftstackrelawd}
\newlength{\leftstackrelbwd}
\def\leftstackrel#1#2{\settowidth{\leftstackrelawd}%
	{${{}^{#1}}$}\settowidth{\leftstackrelbwd}{$#2$}%
	\addtolength{\leftstackrelawd}{-\leftstackrelbwd}%
	\leavevmode\ifthenelse{\lengthtest{\leftstackrelawd>0pt}}%
	{\kern-.5\leftstackrelawd}{}\mathrel{\mathop{#2}\limits^{#1}}}

\newcommand{\mbG}{\bm{G}}
\newcommand{\mbH}{\bm{H}}

\newcommand{\mbX}{\bm{X}}
\newcommand{\mbY}{\bm{Y}}

\newcommand{\mbz}{\bm{z}}

\newcommand{\op}[1]{{\operatorname{#1}}}
\newcommand{\uproman}[1]{\uppercase\expandafter{\romannumeral#1}}
\newcommand{\tr}{\operatorname{tr}}
\newcommand{\vect}{\operatorname{vec}}
\newcommand{\diag}{\operatorname{diag}}

\newcommand{\del}{_{\bm{\delta}}}
\newcommand{\deli}{_{\bm{\delta}_i}}
\newcommand{\eye}{\bm{\op{I}}}
\newcommand{\B}[1]{\bm{#1}}
\newcommand{\inv}{^{-1}}

\newcommand{\h}{^{\op H}}
\newcommand{\T}{^{\op T}}
\newcommand{\siginv}{\frac{1}{\sigma^2}}

\newcommand{\Chat}{\widehat{\B C}}
\newcommand{\chat}{\hat{\B c}}

\newcommand{\GE}{_{\text{GE}}}

\newcommand{\bdelt}{\B\delta}
\newcommand{\Wdel}{\B W\del}
\newcommand{\Cdel}{\B C\del}

\newcommand{\Wha}{\widehat{\B W}}
\newcommand{\dB}{$\operatorname{dB}$}

\usepackage[acronym,shortcuts]{glossaries}
\newacronym{adc}{ADC}{analog-to-digital converter}
\newacronym{aoa}{AOA}{angle of arrival}
\newacronym{aod}{AOD}{angle of direction}
\newacronym{blmmse}{BLMMSE}{Bussgang LMMSE}
\newacronym{bs}{BS}{base station}
\newacronym{ce}{CE}{channel estimation}
\newacronym{cnn}{CNN}{convolutional neural network}
\newacronym{dft}{DFT}{discrete Fourier transform}
\newacronym{em}{EM}{expectation-maximization}
\newacronym{emiht}{EM-IHT}{EM algorithm with IHT}
\newacronym{fe}{FE}{fast estimator}
\newacronym{fft}{FFT}{fast Fourier transform}
\newacronym{flop}{FLOP}{floting point operation}
\newacronym{ge}{GE}{gridded estimator}
\newacronym{iht}{IHT}{iterative hard thresholding}
\newacronym{lmmse}{LMMSE}{linear minimum mean square error}
\newacronym{ls}{LS}{least squares}
\newacronym{mimo}{MIMO}{multiple input multiple output}
\newacronym{ml}{ML}{maximum likelihood}
\newacronym{mmse}{MMSE}{minimum mean squared error}
\newacronym{ms}{MS}{mobile station}
\newacronym{mse}{MSE}{mean squared error}
\newacronym{omp}{OMP}{orthogonal matching pursuit}
\newacronym{relu}{ReLU}{rectified linear unit}
\newacronym{se}{SE}{structured estimator}
\newacronym{simo}{SIMO}{single input multiple output}
\newacronym{snr}{SNR}{signal-to-noise ratio}
\newacronym{ula}{ULA}{uniform linear array}
\newacronym{dl}{DL}{downlink}
\newacronym{ul}{UL}{uplink}
\newacronym{fdd}{FDD}{frequency division duplex}
\newacronym{tdd}{TDD}{time division duplex}
\newacronym{mt}{MT}{mobile terminal}
\newacronym{csi}{CSI}{channel state information}
\newacronym{uma}{UMa}{urban macrocell}
\newacronym{umi}{UMi}{urban microcell}
\newacronym{mpc}{MPC}{multi-path component}
\newacronym{nlos}{NLOS}{non-line of sight}
\newacronym{los}{LOS}{line of sight}
\newacronym{o2i}{O2I}{outdoor-to-indoor}
\newacronym{cs}{CS}{compressive sensing}
\newacronym{cdf}{CDF}{cumulative distribution function}
\newacronym{iqr}{IQR}{interquartile range}

\usepackage{siunitx}
\usepackage[normalem]{ulem}
\usepackage{pgfplots}
\pgfplotsset{compat=1.15}
\usepgfplotslibrary{statistics}

\newcommand{\boxplotheight}{0.625\columnwidth} 
\newcommand{\boxplotwidth}{0.85\columnwidth}

\newcommand{\normalplotheight}{0.7\columnwidth}
\newcommand{\normalplotwidth}{0.8\columnwidth}

\newcommand{\Nrx}{N_{\mathrm{rx}}}
\newcommand{\Ntx}{N_{\mathrm{tx}}}

\newcommand{\lineWidth}{1.0pt}
\tikzset{algorithm1/.style={mark options={solid},color=TUMBeamerBlue, line width=\lineWidth, mark=square, dashed}}

\def\BibTeX{{\rm B\kern-.05em{\sc i\kern-.025em b}\kern-.08em
    T\kern-.1667em\lower.7ex\hbox{E}\kern-.125emX}}

\begin{document}

\title{
Centralized Learning of the Distributed Downlink Channel Estimators in FDD Systems using\\ Uplink Data
\\
\thanks{This work was funded by Huawei Sweden Technologies AB, Lund.}
}

\author{
	\IEEEauthorblockN{Benedikt Fesl, Nurettin Turan, Michael Koller, Michael Joham, and Wolfgang Utschick}
	\IEEEauthorblockA{Professur f\"ur Methoden der Signalverarbeitung, Technische Universit\"at M\"unchen, 80290 Munich, Germany\\
		Email: \{benedikt.fesl, nurettin.turan, michael.koller, joham, utschick\}@tum.de}
}

\maketitle

\begin{abstract}
In this work, we propose a \ac{cnn} based low-complexity approach for \ac{dl} \ac{ce} in \ac{fdd} systems. In contrast to existing work, we use training data which solely stems from the \ac{ul} domain. This allows to learn the \ac{cnn} centralized at the \ac{bs}. After training, the network parameters are offloaded to \acp{mt} within the coverage area of the \ac{bs}. The \acp{mt} can then obtain \ac{csi} of the MIMO channels with the low-complexity \ac{cnn} estimator. This circumvents the necessity of an infeasible amount of feedback, i.e., acquisition of training data at the user, and the offline training phase at each \ac{mt}.  
Numerical results show that the \ac{cnn} which is trained solely based on \ac{ul} data performs equally well as the network trained based on \ac{dl} data. Furthermore, the approach is able to outperform state-of-the-art \ac{ce} algorithms.
\end{abstract}


\IEEEpeerreviewmaketitle

\section{Introduction}\label{sec:intro}
\begin{figure}[b]
\onecolumn
\centering
\copyright This work has been submitted to the IEEE for possible publication. Copyright may be transferred without notice, after which this version may no longer be accessible.

\vspace{-2.05cm}
\twocolumn
\end{figure}

Massive MIMO is a key technology to greatly enhance the wireless communication capacity and throughput due to the increased degrees of freedom \cite{5595728}. To exploit the spatial multiplexing and array gains to their full extent, accurate knowledge of the \ac{csi} is important. A variety of approaches for \ac{ce} concentrate on \ac{tdd} mode, where channel reciprocity is present and the \ac{ul} \ac{csi} can be exploited for the \ac{dl} as well. However, due to the limited coherence time and the calibration error of radio frequency chains, the estimated \ac{csi} at the \ac{bs} may not be accurate for the \ac{dl} \cite{6891254}. Furthermore, in \ac{fdd} mode, more efficient communication with lower latency, improved coverage, and reduced interference can be realized \cite{6888467}. This makes \ac{fdd} an attractive option for future generations of mobile communication.

In \ac{fdd} systems, channel reciprocity does not hold and \ac{dl} \ac{ce} has to be performed at the user who then sends some information regarding the \ac{dl} \ac{csi} via a feedback link to the \ac{bs}.
Due to the large dimensions, this is especially challenging in massive MIMO because of an increasing pilot overhead with increasing number of \ac{bs} antennas, a significantly large feedback size, and limited computational power at the user due to energy efficiency constraints. 
Many approaches for \ac{dl} \ac{ce} in \ac{fdd} systems take structural information like sparsity or low-rank models into account in order to utilize \ac{cs}-based reconstruction techniques \cite{Rao2014} -- \nocite{7174558}\nocite{7355354}\cite{7893736}. However, in many of those works, the feedback overhead is very high and would be infeasible in real systems. Furthermore, the 
\ac{cs}-based approaches do not account for complex statistical correlations that are hard to model nor do they consider structure beyond sparse or low-rank patterns.

Very recently, deep learning based approaches have been used for pilot design, channel estimation, compression, feedback design, or a combination of these topics, e.g., \cite{9037126} --\nocite{9410430} \cite{guo2021canet}. By training the networks with channels from the corresponding scenario, it is possible to learn complex underlying statistical correlations and structures. Consequently, these approaches are generally able to outperform the \ac{cs}-based approaches. However, in none of these works the following crucial aspects are discussed:
\begin{itemize}
    \item Generation of a training dataset in operation mode at the \ac{mt} with measured channels; alternatively, the sharing of a dataset which is stored at the \ac{bs} with the \ac{mt} which causes excessive overhead.
    \item Storage of a training dataset at the \ac{mt}.
    \item Computational complexity and power consumption of the offline training, especially when users move from one cell to another with different environmental structures and a different \ac{bs} location where re-training may be necessary.
\end{itemize}

\subsection{Proposed Learning Strategy}\label{subsec:strategy}
\begin{figure}[t]
 \centering

\resizebox{0.475\textwidth}{!}{%

\tikzset{every picture/.style={line width=0.75pt}} 

\begin{tikzpicture}[x=0.75pt,y=0.75pt,yscale=-1,xscale=1]
	
	\draw    (383,112) .. controls (348.35,132.79) and (333.3,87.91) .. (287.4,87.01) ;
	\draw [shift={(286,87)}, rotate = 360] [color={rgb, 255:red, 0; green, 0; blue, 0 }  ][line width=0.75]    (10.93,-3.29) .. controls (6.95,-1.4) and (3.31,-0.3) .. (0,0) .. controls (3.31,0.3) and (6.95,1.4) .. (10.93,3.29)   ;
	\draw  [fill={rgb, 255:red, 74; green, 144; blue, 226 }  ,fill opacity=0.3 ] (232,170) .. controls (232,165.58) and (235.58,162) .. (240,162) -- (270,162) .. controls (274.42,162) and (278,165.58) .. (278,170) -- (278,194) .. controls (278,198.42) and (274.42,202) .. (270,202) -- (240,202) .. controls (235.58,202) and (232,198.42) .. (232,194) -- cycle ;
	\draw    (255,117) -- (255,148) ;
	\draw [shift={(255,150)}, rotate = 270] [color={rgb, 255:red, 0; green, 0; blue, 0 }  ][line width=0.75]    (10.93,-3.29) .. controls (6.95,-1.4) and (3.31,-0.3) .. (0,0) .. controls (3.31,0.3) and (6.95,1.4) .. (10.93,3.29)   ;
	\draw  [fill={rgb, 255:red, 245; green, 166; blue, 35 }  ,fill opacity=0.3 ] (79,73) .. controls (79,68.58) and (82.58,65) .. (87,65) -- (117,65) .. controls (121.42,65) and (125,68.58) .. (125,73) -- (125,97) .. controls (125,101.42) and (121.42,105) .. (117,105) -- (87,105) .. controls (82.58,105) and (79,101.42) .. (79,97) -- cycle ;
	\draw    (222,87) -- (140,87) ;
	\draw [shift={(138,87)}, rotate = 360] [color={rgb, 255:red, 0; green, 0; blue, 0 }  ][line width=0.75]    (10.93,-3.29) .. controls (6.95,-1.4) and (3.31,-0.3) .. (0,0) .. controls (3.31,0.3) and (6.95,1.4) .. (10.93,3.29)   ;
	\draw  [color={rgb, 255:red, 128; green, 128; blue, 128 }  ,draw opacity=1 ] (354,221.4) .. controls (354,217.87) and (356.87,215) .. (360.4,215) -- (379.6,215) .. controls (383.13,215) and (386,217.87) .. (386,221.4) -- (386,259.6) .. controls (386,263.13) and (383.13,266) .. (379.6,266) -- (360.4,266) .. controls (356.87,266) and (354,263.13) .. (354,259.6) -- cycle ;
	\draw  [color={rgb, 255:red, 128; green, 128; blue, 128 }  ,draw opacity=1 ] (357,221.4) -- (383,221.4) -- (383,259.6) -- (357,259.6) -- cycle ;
	\draw  [color={rgb, 255:red, 128; green, 128; blue, 128 }  ,draw opacity=1 ] (405,78.4) .. controls (405,74.87) and (407.87,72) .. (411.4,72) -- (430.6,72) .. controls (434.13,72) and (437,74.87) .. (437,78.4) -- (437,116.6) .. controls (437,120.13) and (434.13,123) .. (430.6,123) -- (411.4,123) .. controls (407.87,123) and (405,120.13) .. (405,116.6) -- cycle ;
	\draw  [color={rgb, 255:red, 128; green, 128; blue, 128 }  ,draw opacity=1 ] (408,78.4) -- (434,78.4) -- (434,116.6) -- (408,116.6) -- cycle ;
	\draw  [color={rgb, 255:red, 128; green, 128; blue, 128 }  ,draw opacity=1 ] (454,91.4) .. controls (454,87.87) and (456.87,85) .. (460.4,85) -- (479.6,85) .. controls (483.13,85) and (486,87.87) .. (486,91.4) -- (486,129.6) .. controls (486,133.13) and (483.13,136) .. (479.6,136) -- (460.4,136) .. controls (456.87,136) and (454,133.13) .. (454,129.6) -- cycle ;
	\draw  [color={rgb, 255:red, 128; green, 128; blue, 128 }  ,draw opacity=1 ] (457,91.4) -- (483,91.4) -- (483,129.6) -- (457,129.6) -- cycle ;
	\draw  [color={rgb, 255:red, 128; green, 128; blue, 128 }  ,draw opacity=1 ] (419,139.4) .. controls (419,135.87) and (421.87,133) .. (425.4,133) -- (444.6,133) .. controls (448.13,133) and (451,135.87) .. (451,139.4) -- (451,177.6) .. controls (451,181.13) and (448.13,184) .. (444.6,184) -- (425.4,184) .. controls (421.87,184) and (419,181.13) .. (419,177.6) -- cycle ;
	\draw  [color={rgb, 255:red, 128; green, 128; blue, 128 }  ,draw opacity=1 ] (422,139.4) -- (448,139.4) -- (448,177.6) -- (422,177.6) -- cycle ;
	\draw  [fill={rgb, 255:red, 126; green, 211; blue, 33 }  ,fill opacity=0.3 ] (230,74) .. controls (230,69.58) and (233.58,66) .. (238,66) -- (268,66) .. controls (272.42,66) and (276,69.58) .. (276,74) -- (276,98) .. controls (276,102.42) and (272.42,106) .. (268,106) -- (238,106) .. controls (233.58,106) and (230,102.42) .. (230,98) -- cycle ;
	\draw    (102,113) .. controls (102,129.66) and (213.41,141.52) .. (240.45,148.57) ;
	\draw [shift={(242,149)}, rotate = 196.26] [color={rgb, 255:red, 0; green, 0; blue, 0 }  ][line width=0.75]    (10.93,-3.29) .. controls (6.95,-1.4) and (3.31,-0.3) .. (0,0) .. controls (3.31,0.3) and (6.95,1.4) .. (10.93,3.29)   ;
	\draw  [fill={rgb, 255:red, 74; green, 144; blue, 226 }  ,fill opacity=0.3 ] (359,229) .. controls (359,227.34) and (360.34,226) .. (362,226) -- (378,226) .. controls (379.66,226) and (381,227.34) .. (381,229) -- (381,238) .. controls (381,239.66) and (379.66,241) .. (378,241) -- (362,241) .. controls (360.34,241) and (359,239.66) .. (359,238) -- cycle ;
	\draw    (381,234) -- (406.33,234.31) ;
	\draw [shift={(408.33,234.33)}, rotate = 180.7] [color={rgb, 255:red, 0; green, 0; blue, 0 }  ][line width=0.75]    (10.93,-3.29) .. controls (6.95,-1.4) and (3.31,-0.3) .. (0,0) .. controls (3.31,0.3) and (6.95,1.4) .. (10.93,3.29)   ;
	\draw  [color={rgb, 255:red, 128; green, 128; blue, 128 }  ,draw opacity=1 ] (410.5,85.7) .. controls (410.5,83.93) and (411.93,82.5) .. (413.7,82.5) -- (428.3,82.5) .. controls (430.07,82.5) and (431.5,83.93) .. (431.5,85.7) -- (431.5,95.3) .. controls (431.5,97.07) and (430.07,98.5) .. (428.3,98.5) -- (413.7,98.5) .. controls (411.93,98.5) and (410.5,97.07) .. (410.5,95.3) -- cycle ;
	\draw  [color={rgb, 255:red, 128; green, 128; blue, 128 }  ,draw opacity=1 ] (459.5,98.7) .. controls (459.5,96.93) and (460.93,95.5) .. (462.7,95.5) -- (477.3,95.5) .. controls (479.07,95.5) and (480.5,96.93) .. (480.5,98.7) -- (480.5,108.3) .. controls (480.5,110.07) and (479.07,111.5) .. (477.3,111.5) -- (462.7,111.5) .. controls (460.93,111.5) and (459.5,110.07) .. (459.5,108.3) -- cycle ;
	\draw  [color={rgb, 255:red, 128; green, 128; blue, 128 }  ,draw opacity=1 ] (424.5,147.7) .. controls (424.5,145.93) and (425.93,144.5) .. (427.7,144.5) -- (442.3,144.5) .. controls (444.07,144.5) and (445.5,145.93) .. (445.5,147.7) -- (445.5,157.3) .. controls (445.5,159.07) and (444.07,160.5) .. (442.3,160.5) -- (427.7,160.5) .. controls (425.93,160.5) and (424.5,159.07) .. (424.5,157.3) -- cycle ;
	\draw    (283,181) .. controls (326.5,182.5) and (346,208) .. (362,226) ;
	\draw    (378,241) -- (386.14,251.05) ;
	\draw [shift={(387.4,252.6)}, rotate = 230.98] [color={rgb, 255:red, 0; green, 0; blue, 0 }  ][line width=0.75]    (10.93,-3.29) .. controls (6.95,-1.4) and (3.31,-0.3) .. (0,0) .. controls (3.31,0.3) and (6.95,1.4) .. (10.93,3.29)   ;
	\draw    (326.33,233.33) -- (356.67,233.65) ;
	\draw [shift={(358.67,233.67)}, rotate = 180.59] [color={rgb, 255:red, 0; green, 0; blue, 0 }  ][line width=0.75]    (10.93,-3.29) .. controls (6.95,-1.4) and (3.31,-0.3) .. (0,0) .. controls (3.31,0.3) and (6.95,1.4) .. (10.93,3.29)   ;
	
	\draw (426,49) node [anchor=north west][inner sep=0.75pt]   [align=left] {MT};
	\draw (325,123) node [anchor=north west][inner sep=0.75pt]   [align=left] 
	{$\mathcal{H}^{\text{UL}}$};
	\draw (180,120) node [anchor=north west][inner sep=0.75pt]   [align=left] {learn CNN};
	\draw (238,174) node [anchor=north west][inner sep=0.75pt]   [align=left] {CNN};
	\draw (296,63) node [anchor=north west][inner sep=0.75pt]   [align=left] {collect UL data};
	\draw (95,77) node [anchor=north west][inner sep=0.75pt]   [align=left] {$\mathcal{D}$};
	\draw (140,65) node [anchor=north west][inner sep=0.75pt]   [align=left] 
	{$\tilde{\B H}_{i,\text{DL}} \leftarrow \B H_{i,\text{UL}}\T$\\ \vspace{-0.25cm} \\~~~emulate DL};
	\draw (242,78) node [anchor=north west][inner sep=0.75pt]   [align=left] {BS};
	\draw (361,231.6) node [anchor=north west][inner sep=0.75pt]  [font=\tiny] [align=left] {CNN};
	\draw (412.67,228.67) node [anchor=north west][inner sep=0.75pt]   [align=left] {estimated channel ${\mbH}$};
	\draw (390.67,200.67) node [anchor=north west][inner sep=0.75pt]   [align=left] {MT};
	\draw (292.79,155.97) node [anchor=north west][inner sep=0.75pt]  [rotate=-27.38] [align=left] {share parameters};
	\draw (240.17,228.67) node [anchor=north west][inner sep=0.75pt]   [align=left] {observation $\mbY$};

\end{tikzpicture}
}
    \caption{The proposed approach of learning the CNN \ac{ce} at the BS, which is then subsequently offloaded to the \ac{mt}.}
    \label{fig:my_label}
    \vspace{-0.4cm}
\end{figure}
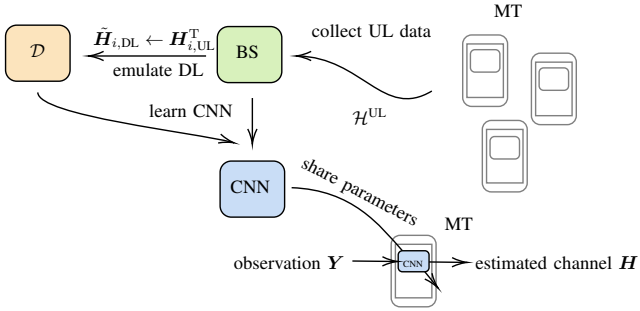

In order to prevent these described problems of learning-based approaches, we propose a learning strategy where the network is trained solely based on \ac{ul} channel data directly at the \ac{bs} by emulating the \ac{dl} system with transposed \ac{ul} channels. Only the network parameters, i.e., the weights and biases of the neural network, have to be shared with the user in an initialization stage, e.g., when a \ac{mt} enters the coverage area of the \ac{bs}. Fig. \ref{fig:my_label} summarizes the procedure:
\begin{enumerate}
    \item Construction of a training dataset $\mathcal{H}^{\text{UL}}$ from \ac{ul} channels at the \ac{bs}.
    \item Transposition of the channels in $\mathcal{H}^{\text{UL}}$ and emulation of the \ac{dl} system to train \ac{cnn} at the \ac{bs}.
    \item Offloading of the \ac{cnn} parameters to the \acp{mt}.
    \item \ac{ce} at the \ac{mt} with low-complexity \ac{cnn}.
\end{enumerate}

A further possibility is a signaling approach where \acp{mt} in a certain sector of the coverage area can be updated with adjusted network parameters that are trained offline beforehand, e.g., if a \ac{mt} changes from a \ac{los} to a \ac{nlos} channel, for example, by moving inside or behind a building, and therefore has different channel statistics. In other words, because the network is trained centralized at the \ac{bs}, the opportunity to provide \acp{mt} in different sectors of the coverage area with different network parameters is given, due to available long-term statistics which can be inferred from the \ac{ul} domain. 

The motivation for this procedure stems from a recently developed \ac{ul}-\ac{dl} conjecture, stating that the distributions of \ac{ul} and \ac{dl} channels are very similar, although the instantaneous channels between \ac{ul} and \ac{dl} are highly different \cite{utschick2021learning}, \cite{rizzello2021learning}. This follows from the fact that a frequency shift, which is present between \ac{ul} and \ac{dl} in \ac{fdd} systems, can be alternatively described by a change in the path length, or by a change of the antenna positioning, at least for narrowband signals. Consequently, for a certain propagation scenario it can be expected that there exist different pairs of antenna positions and carrier frequencies with identical \ac{csi} \cite{utschick2021learning}. In \cite{turan2021unsupervised}, this \ac{ul}-\ac{dl} conjecture is exploited for adaptive codebook construction and feedback generation.

In general, this learning strategy is applicable to any learning-based approach which is trained on channel data. In this work, we apply it to the recently proposed low-complexity \ac{cnn} estimator for MIMO channels \cite{mimocnn} which is based on the original approach from \cite{8272484} and learns a model-based \ac{cnn} architecture, motivated by the conditional mean estimator. Numerical results demonstrate that the \ac{cnn} estimator trained on \ac{ul} data performs equally well in terms of \ac{mse} as the \ac{cnn} estimator trained on \ac{dl} data. Moreover, the proposed approach is able to outperform state-of-the-art \ac{ce} approaches in different scenarios.

\textbf{Notation:} The transpose and conjugate-transpose of a vector $\B x$ are denoted by $\B x\T$ and $\B x\h$.
The $n\times n$ identity matrix and the $n\times 1$ all-ones vector are denoted by $\eye_n$ and $\B 1_n$, respectively. The column-wise vectorization and the trace of a matrix $\B X$ are denoted by $\vect(\B X)$ and $\tr(\B X)$ and the Kronecker product by $ \otimes$. A diagonal matrix with diagonal $\B x$ is given by $\diag(\B x)$. 

\section{System Model}
We consider a \ac{dl} scenario, where $N_P$ pilot signals are transmitted. At the \ac{bs} and the \ac{mt} \acp{ula} with $\Ntx$ and $\Nrx$ antennas are deployed, respectively.
The channel is assumed to be frequency-flat with block-fading such that we get independent observations in each coherence interval.
We investigate a single-snapshot scenario, i.e., the coherence interval of the covariance matrix and of the channel is identical. The received signal is $\B Y = \B H \B X^\prime + \B  Z$
with the channel matrix $\B H\in\mathbb{C}^{\Nrx\times \Ntx}$ and the pilot matrix $\B X^\prime\in\mathbb{C}^{\Ntx\times N_P}$.
After vectorization we get
\begin{equation}
	\B y = \B X\B h + \B z \in 	\mathbb{C}^{\Nrx N_P},
\end{equation}
with $\B X = \B X^{\prime,\op T} \otimes \eye_{\Nrx}$ and $\B h = \vect(\B H)\hspace{-0.05cm}\in\hspace{-0.05cm}\mathbb{C}^{\Ntx\Nrx}$. Note that we use the channel matrix $\B H$ and its vectorized expression $\B h$ interchangeably in the following for the ease of notation.
We choose a scaled \ac{dft} matrix $\B X^\prime = \frac{1}{\sqrt{\Ntx}}\B F$ as pilot matrix, where $\B F$ is the \ac{dft} matrix \cite{Biguesh2006}.
The noise vector is assumed to be distributed as $\B z \hspace{-0.05cm}\sim \hspace{-0.05cm}\mathcal{N}_{\mathbb{C}}(\B 0,\sigma^2\eye_{\Nrx N_P})$.

\section{Channel Model  and Data Generation}\label{sec:datagen}
We consider an \ac{fdd} system and assume a frequency gap of $\SI{200}{\mega\hertz}$ between \ac{ul} and \ac{dl}.
Version $2.2$ of the QuaDRiGa channel simulator \cite{QuaDRiGa1, QuaDRiGa2} is used to generate \ac{csi} for the \ac{ul} and \ac{dl} scenarios.
We simulate two \ac{uma} single carrier scenarios: one with $(\Ntx, \Nrx) = (8, 4)$ and one with $ (\Ntx, \Nrx) = (64, 4)$.
In both cases, the \ac{ul} carrier frequency is $\SI{2.53}{\giga\hertz}$ and the \ac{dl} carrier frequency is $\SI{2.73}{\giga\hertz}$. 
The \ac{bs} is equipped with a \ac{ula} with ``3GPP-3D'' antennas, and the \ac{mt} consists of a \ac{ula} assuming ``omni-directional'' antennas. 
The \ac{bs} is placed at a height of $\SI{25}{\meter}$ and covers a sector of $\SI{120}{\degree}$, where the minimum distance of the \ac{mt} location to the \ac{bs} is $\SI{35}{\meter}$ and the maximum distance to the \ac{bs} is $\SI{500}{\meter}$. 
In $80\%$ of the cases, the \ac{mt} is located indoors at different floor levels, and in the case of outdoor locations the \ac{mt}'s height is $\SI{1.5}{\meter}$ in accordance with \cite{3gpp_uma}. Simulation results were also carried out for different center frequencies and for the \ac{umi} scenario, but displayed qualitatively the same results and are therefore not shown here.

As outlined in \cite{QuaDRiGa1}, many parameters such as path-loss, delay, and angular spreads, path-powers for each subpath, and antenna patterns are different in the \ac{ul} and \ac{dl} domain. 
However, the following parameters are identical in the \ac{ul} and \ac{dl} domain: \ac{bs} location and the \ac{mt} locations, propagation cluster delays and angles for each \ac{mpc}, and the spatial consistency of the large scale fading parameters. 
QuaDRiGa models MIMO channels as
\begin{equation} 
\mbH = \sum_{\ell=1}^{L} \mbG_{\ell} e^{-2\pi j f_c \tau_{\ell}}, 
\end{equation}
where $\ell$ is the path number, and the number of \acp{mpc} $L$ depends on whether there is \ac{los}, \ac{nlos}, or \ac{o2i} propagation: $L_\text{LOS} = 37$, $L_\text{NLOS} = 61$, or $L_\text{O2I} = 37$.
The carrier frequency is denoted by $f_c$ and the $\ell$-th path delay by $\tau_{\ell}$. 
The MIMO coefficient matrix $\mbG_{\ell}$ consists of one complex entry for each antenna pair, which comprises the attenuation of a path, the antenna radiation pattern weighting, and the polarization \cite{Kurras}.
Each channel is normalized using its path gain as $\mbH = 10^{-0.05\text{PG[dB]}}\mbH_{raw}$ according to \cite{QuaDRiGa2}.

We generate datasets with $31\times 10^3$ channels for both the \ac{ul} and \ac{dl} of each scenario, where $1000$ channels are used to calculate a global sample covariance matrix 
\begin{equation}
    \Chat_{\text{glob}} = \frac{1}{1000}\sum_{i=1}^{1000} \B h_i\B h_i\h,
    \label{eq:samplecov}
\end{equation}
for both \ac{ul} and \ac{dl}, respectively. The channels which are used for the calculation of the sample covariance matrix are not further used for training or evaluation. 
The training dataset contains $20\times 10^3$ channels, whereas the performance of the different approaches are averaged over $10\times 10^3$ channels in the test dataset.
For each dataset of \ac{ul} and \ac{dl}, the channels are normalized, such that it holds: $\op E [||\B h ||^2] = \Ntx\Nrx$.
Accordingly, we define the \ac{snr} as $\text{SNR} = {E[\|\mbX \|^2_F]}/{E[\|\mbz \|^2_2]}=1/\sigma^2$. 
This leads to the following pairs of \ac{ul} and \ac{dl} datasets for the two considered scenarios:
\begin{align}
(\mathcal{H}_{4\times8}^{\text{UL}}, \mathcal{H}_{4\times8}^{\text{DL}}) \ \text{and} \ (\mathcal{H}_{4\times64}^{\text{UL}}, \mathcal{H}_{4\times64}^{\text{DL}}).
\end{align}

The \ac{ul} channels have a dimension of $8\times4$ or $64\times4$ depending on the scenario. The sets $\mathcal{H}_{4\times8}^{\text{UL}}$ and $\mathcal{H}_{4\times64}^{\text{UL}}$ contain transposed versions of the respective channels, i.e, with dimensions $4\times8$ or $4\times64$.

\section{Revision of the CNN Estimator}\label{sec:mimocnn}
In the following, we give a short revision of the \ac{cnn} estimator from \cite{mimocnn}.
We assume that for a given variable $\bdelt$, the channels are conditionally Gaussian distributed as $\B h|\bdelt\sim\mathcal{CN}(\B 0, \Cdel)$. The vector $\bdelt$ contains prior information such as angles of arrival or path gains and follows an unknown distribution $\bdelt\sim p(\bdelt)$.
Assuming knowledge of the prior parameters $\bdelt$, the conditional \ac{mmse} estimate of $\B h$ from $\B y$ would read as
\begin{align}
	\op E[\B h|\B y,\bdelt] &= \op E[\B h\B y\h|\bdelt]\op E[\B y\B y\h|\bdelt]\inv \B y\\
	&=\Cdel\B X\h(\B X\Cdel\B X\h + \sigma^2\eye_{\Nrx N_P})\inv\B y \label{eq:cond_mmse}
	= \Wdel\B y,
\end{align} 
which depends linearly on the observations. However, as the parameters $\bdelt$ are unknown in general, the law of total expectation is used to compute \cite{8272484}
\begin{align}
	\hat{\B h} &= \op E[\B h|\B y] = \op E[\op E[\B h|\B y,\bdelt]|\B y] 
	= \op E[\Wdel\B y|\B y] 
	= \Wha_\star(\B y)\B y.
\end{align} 
Thus, to obtain $\hat{\B h}$, the conditional mean $\Wha_\star$ of the \ac{mmse} 
filter $\Wdel$ has to be determined, which nonlinearly depends on $\B y$.

Bayes' theorem is used to state the \ac{mmse} filter as \cite{8272484}
\begin{equation}
	\label{eq:Wstar}
	\Wha_\star = \int p(\bdelt|\B y) \Wdel \op d\bdelt = \frac{\int p(\B y|\bdelt)\Wdel p(\bdelt)\op d\bdelt }{\int p(\B y|\bdelt) p(\bdelt)\op d\bdelt }.
\end{equation}
As shown in \cite{mimocnn}, due to the Gaussian assumption of $\B y$ given $\bdelt$, the \ac{mmse} filter can be written~as
\begin{equation}\label{eq:mmse}
	\Wha(\Chat) = \frac{\int \exp(\tr(\B X\Wdel\Chat) + \log|\eye-\B X\Wdel|)\Wdel p(\bdelt)\op d\bdelt}{\int \exp(\tr(\B X\Wdel\Chat) + \log|\eye-\B X\Wdel|) p(\bdelt)\op d \bdelt},
\end{equation}	
with the scaled single-shot sample covariance matrix $	\Chat =\siginv\B y\B y\h$ as input.
The \ac{mmse} filter $\Wha_\star$ depends on the observations through $\Chat$ and is thus a nonlinear filter. For an arbitrary distribution $p(\bdelt)$, the \ac{mmse} filter in \eqref{eq:mmse} is not computable. To overcome this problem, the prior distribution is assumed to be discrete and uniform on a grid $\{\bdelt_i:i=1,\dots,P \}$ with $p(\bdelt_i) = 1 / P$ for all $i=1,...,P$, cf. \cite{mimocnn}.
With this assumption, the \ac{mmse} estimator is~evaluated~as
\begin{equation}\label{eq:ge}
	\Wha\GE(\Chat)=  \frac{1/P\sum_{i=1}^P \exp(\tr(\B X\B W\deli\Chat) +b_i)\B W\deli}{1/P\sum_{i=1}^P \exp(\tr(\B X\B W\deli\Chat) + b_i)},
\end{equation}
with $b_i = \log|\eye-\B X\B W\deli|$. This approximates $ \Wha(\Chat) $, where the approximation error decreases with increasing $P$ \cite{8272484}. 

The complexity of the estimator can be reduced by exploiting common structure of the covariance matrices, i.e., it is assumed that the channel covariance matrix is a block-circulant matrix with circulant blocks, which only holds asymptotically for large-scale systems \cite{mimocnn}. As such a matrix is decomposed by a two-dimensional \ac{dft} matrix $\B Q$, i.e., 
$\Cdel = \B Q \h \diag(\B c_{\bdelt} ) \B Q$, the \ac{mmse} filter from \eqref{eq:cond_mmse} can be decomposed as $\B W\del = \B Q\h \diag(\B w\del) \B Q$. 
With this assumption, the \ac{mmse} filter from \eqref{eq:ge} is approximately
\begin{align}\label{eq:ass2}
	\Wha(\chat) \approx \B Q\h\diag\left(\B w(\chat)
	\right)\B Q\B X\h ,
\end{align}
with the definitions
\begin{align}
	\B w(\chat) &= \B A\frac{\exp(\B A\T\chat + \B b)}{\B 1_P\T\exp(\B A\T\chat + \B b)},\label{eq:SE_element}\\
	\B A &=[\B w_{\bdelt_1}, \dots, \B w_{\bdelt_P}]\in	\mathbb{C}^{\Nrx\Ntx\times P}, \label{eq:ASE}\\
	\chat&= \sigma^{-2}|\B Q\B X\h\B y|^2\in \mathbb{C}^{\Nrx\Ntx} ,\label{eq:chat}
\end{align}
and $\B b = [b_{\bdelt_1},...,b_{\bdelt_P}]\T$ as shown in \cite[Appendix C]{mimocnn}.

The complexity is further reduced by assuming $\B A$ to be a block-circulant matrix with circulant blocks, which is valid for specific scenarios, i.e., a single propagation cluster, cf. \cite{mimocnn}. Following this assumption, there exists a $\B w_0$ such that $\B A = \B Q\h \diag(\B Q\B w_0) \B Q$.
This is equal to constructing $\B A$ by a two-dimensional circular convolution with $\B w_0$ as convolution kernel and we can write $\B A \B x = \B w_0 \star \B x.$
The estimator containing all three assumptions can then be written as
\begin{equation}\label{eq:fe}
	\Wha(\hat{\B c}) =
	\B Q\h\operatorname{diag}\left(\B w_0 \star \phi(\tilde{\B w}_0\star\chat+ \B b)
	\right)\B Q\B X\h ,
\end{equation}
where $\phi$ is the softmax function and $\tilde{\B w}_0$ contains the entries of $\B w_0$ in reversed order.

These assumptions for the derivation may only rarely be fulfilled in real scenarios. Therefore, the estimator from \eqref{eq:fe} is interpreted as a \ac{cnn} with two two-dimensional convolution layers, which implements a function from the set
\begin{equation}
	\mathcal{W}_{\text{CNN}} = \{\B x \mapsto \B a^{(2)} \star \psi (\B a^{(1)} \star \B x + \B b^{(1)}) + \B b^{(2)}
	\},
\end{equation}
where $\B a^{(l)}, \B b^{(l)}\in\mathbb{R}^{\Nrx\Ntx} , l=1,2$ are the parameters which are learned during training from samples $(\B y_i, \B h_i)$,
Although the softmax function $\phi$ is identified as the activation in \eqref{eq:fe}, the activation is relaxed to be a different function $\psi$, e.g., the well-known \ac{relu}.
The optimal \ac{cnn} estimator is the function which minimizes the MSE, i.e.,
\begin{equation}
	\hat{\B w}_{\text{CNN}} = \underset{ \hat{\B w}(\cdot)\in\mathcal{W}_{\text{CNN}}}{\arg\min}~\op E[||\B h - \B Q\h\operatorname{diag}(\hat{\B w}(\chat))\B Q\B X\h \B y||_2^2].
\end{equation}

Due to the possibility to apply \ac{fft} for the two-dimensional circular convolution and the pilot matrix, the overall complexity of the estimator is only~$\mathcal{O}(\Ntx\Nrx\log(\Nrx\Ntx))$.

\section{Baseline Algorithms}
In the QuaDRiGa software, the channel covariance matrix for each channel realization is not available. Therefore, we do not have perfect knowledge of the second-order channel statistics in order to evaluate the optimal \ac{lmmse} estimator. However, to have a reasonable baseline, we evaluate the \ac{lmmse} with the sampled global covariance matrix $\Chat_{\text{glob}}$, cf. \eqref{eq:samplecov}. The estimate then reads as
\begin{equation}
    \hat{\B h}_{\text{LMMSE}} = \Chat_{\text{glob}}\B X\h (\B X \Chat_{\text{glob}}\B X\h + \sigma^2\eye_{\Nrx N_P})\inv \B y.
\end{equation}
Note that this clearly is a suboptimal approach as the global covariance matrix is different from the conditional covariance matrix $\Cdel$ which takes account of the prior $\bdelt$. To further evaluate the \ac{ul}-\ac{dl} conjecture described above, we evaluate the \ac{lmmse} for both, the sampled global covariance matrix of \ac{ul} and \ac{dl} channels.

In massive MIMO systems, the channel covariance matrices have a low numerical rank \cite{7499112}, and therefore, the channel vector $\B h$ can be approximated by a sparse vector $\B t$ for a given dictionary $\B D$, i.e.,
$\B h \approx \B D\B t$.
A reasonable choice for the dictionary is $\B D = \B D_{\text{rx}} \otimes \B D_{\text{tx}}$, where $ \B D_{\text{rx}}$ and $ \B D_{\text{tx}}$ are oversampled \ac{dft} matrices \cite{7178503}. 
This choice is motivated by the fact that the \ac{dft} matrix maps to the sparse angular domain in case of a \ac{ula}.
The vector $\B t$ can then be found by solving a sparse approximation problem, for which the \ac{omp} algorithm is suitable \cite{681706}.
As the sparsity order is unknown, we use a genie-aided upper bound in our simulations, which decides about the sparsity level with the given exact channel realization. 

Another low-complexity \ac{ce} algorithm is \ac{ml} estimation of the structured covariance matrix \cite{7051818}, \cite{1456695}. The estimate with the block-circular assumption for the covariance matrix is
\begin{equation*}
	\hat{\B h} = \B Q\h\diag(\B c_{\bdelt}^{\text{ML}})(NU\inv\diag(\B c_{\bdelt}^{\text{ML}}) + \sigma^2\eye_{SU})\inv\tilde{\B Q}\B y
\end{equation*}
with $\tilde{\B Q} = \B Q\B X\h$. 
The eigenvalues of $\Cdel$ are estimated as
$\B c_{\bdelt}^{\text{ML}} = [\B s - \sigma^2\B 1]_+$, where $\B s = |\B Q \B X\h \B y|^2$ and the $i$th element of $[\B x]_+$ is $\max( x_i,0)$, cf. \cite{8272484}.
We further show the \ac{ls} solution which minimizes the $\ell_2$ norm $||\B y - \B X\B h||_2$ \cite{1597555}.

\section{Simulation Results}
\input{plots.tex}

We depict numerical results where we simulated the proposed learning strategy described in Subsection \ref{subsec:strategy} with data that is generated with the QuaDRiGa software, cf. Section \ref{sec:datagen}. 

We evaluate the described \ac{cnn} estimator from Section \ref{sec:mimocnn} which is trained solely on \ac{ul} data and has never seen \ac{dl} channels in the training phase for two different activation functions: \ac{relu} and softmax. We refer to these results as "ReLU UL" and "softmax UL", respectively. Note that the evaluation data always stem from actual \ac{dl} channels. For direct comparison, we depict the same \ac{cnn} estimator which is trained on \ac{dl} channels directly, referred to as "ReLU DL" and "softmax DL", respectively. For verification that the proposed \ac{cnn} estimator indeed learns structural features of the chosen scenario, we show results where the input channels for the \ac{cnn} in the training phase are purely i.i.d. Gaussian channels, without any structural properties that occur in the \ac{uma} scenario. We refer to this results as "ReLU Gauss" and "softmax Gauss", respectively. Lastly, we evaluate the LMMSE estimator which uses the sampled global covariance matrix from \eqref{eq:samplecov}, sampled from \ac{dl} channels ("LMMSE DL") and \ac{ul} channels ("LMMSE UL"), respectively.

In Fig. \ref{fig:snr_mixed}, we compare the different estimators for different \acp{snr} ranging from $-15$\dB ~to $20$\dB ~for a $4\times 8$ (top) and $4\times 64$ (bottom) MIMO setup.
First of all, one can observe that the \ac{cnn} estimator performs equally well when being trained on either data which stem from the \ac{dl} or \ac{ul}. This can be observed for both setups and therefore validates the \ac{ul}-\ac{dl} conjecture stated in \cite{utschick2021learning} and \cite{rizzello2021learning}. One can further observe a performance gap to the \ac{cnn} estimator trained on i.i.d. Gaussian channels. This indicates that the \ac{cnn} estimator is able to adapt to the scenario by inferring structural properties of the data that is provided in the training phase. Since the \ac{ul} and \ac{dl} data stem from the same scenario, it seems that common features are shared between both domains. Furthermore, the performance gap as compared to training with Gaussian data increases for the higher-dimensional setup (bottom), which may be due to more structure within the data in larger setups, e.g., due to antenna correlations.

The same argumentation holds for the comparison with the LMMSE, either based on \ac{ul} or \ac{dl} channels. Due to the global sampling of the covariance matrix, there is no structural information that is exploited by the estimator as this is averaged out (the $\ h_i$ in \eqref{eq:samplecov} are found all over the cell).
If we further compare the \ac{cnn} estimator with the baseline approaches \ac{ls}, \ac{ml}, and genie \ac{omp}, the superiority of the learning-based approach becomes clear. Especially the \ac{cs}-based approach \ac{omp} is not able to compete with the proposed learning-based approach. This may be caused by the difficulty of the \ac{uma} scenario, in which many channels are \ac{nlos} channels with many subpaths, which again are constructed by the superposition of many micropaths. In such scenarios, the underlying sparsity may either not be given, or it can not be exploited by the \ac{omp} algorithm with the given dictionary.

After we have seen that the \ac{cnn} estimator performs equally well on average, independent of whether the training data stems from \ac{ul} or \ac{dl}, we now want to investigate this more precisely by looking at the naturally occurring variations of the estimation performance over many samples in terms of \ac{mse} for a fixed \ac{snr} of $5$\dB. This variation is caused by the difference of the channel samples, e.g., by being \ac{los} or \ac{nlos} channels, and therefore being easier or harder to estimate in general. In Fig. \ref{fig:boxplot}, these variations are evaluated in form of a boxplot (top) and in form of their \acp{cdf} (bottom). For the boxplot, the lower and upper box edges indicate the first and third quartile, respectively, and the middle line the median (50th percentile). For convenience, we also show the mean values (dots). From the lower (upper) box edges, a distance of 1.5 times the \ac{iqr} is measured where a whisker is drawn up to the lowest (largest) observed point from the dataset that falls within this distance. All other observed points are plotted as outliers. 

It can be observed that the boxplots for the estimators based on \ac{ul} and the ones based on \ac{dl} data are very similar and that their \acp{cdf} are lying on top of each other. This means that the estimators are not only performing equally well on average, it can be expected that the estimators also show very similar performance on instantaneous channel realizations. Here, it is important to keep in mind that the test set on which the plots are based are the same and stem from \ac{dl} channels.

We can see further that the \ac{cnn} estimator with \ac{relu} not only shows the best performance on average, but its upper whisker is also bounded tightly at the lowest \ac{mse} compared to the baseline algorithms. The performance of the genie \ac{omp}, however, strongly depends on the instantaneous channel realizations and therefore shows a huge variation in terms of \ac{mse}. When looking at the \ac{cdf}, the genie \ac{omp} can achieve very good performance with low probability, but fails to achieve a certain error bound with high probability in such a difficult scenario.

In addition to the already shown scenario with mixed \ac{nlos} and \ac{los} channels, we now evaluate a \ac{uma} scenario with pure \ac{los} channels. The \ac{los} channels have a single dominant subpath and are therefore more easily sparsly representable. In addition to the approaches we already discussed before, we depict the \ac{cnn} estimator which is trained on the dataset from before with mixed \ac{nlos} and \ac{los} channels, but is evaluated on the pure \ac{los}, referred to as "ReLU mixed" and "softmax mixed". It can be seen that especially the genie \ac{omp} approach benefits from the pure \ac{los} scenario, especially for low to medium \ac{snr} values. However, the learning-based approaches outperform the \ac{cs} approach also in this scenario for higher \ac{snr} values. Besides the observation that the \ac{ul}-\ac{dl} conjecture also holds for the \ac{los} scenario, it conveys that the \ac{cnn} approach with \ac{relu} activation trained on mixed \ac{nlos} and \ac{los} channels almost achieves the same performance as the one that is trained purely on \ac{los} channels. This makes the great generalization ability of the learning-based \ac{cnn} approach visible. The \ac{ls} and LMMSE approach are not able to exploit the structural simplifications of the scenario and therefore show no gain compared to the mixed \ac{nlos} and \ac{los} scenario. 

\section{Conclusion}
In this work, we proposed a novel signaling approach for centralized learning of the distributed \ac{dl} channel estimators in \ac{fdd} systems. This is based on the observation that, regardless of the difference of the instantaneous \ac{ul} and \ac{dl} channels, the \ac{ul} dataset can be used for training the \ac{dl} channel estimator due to the similar distributions of \ac{ul} and \ac{dl} channels. 
This again illustrates nicely the fact that neural networks are not dependent on specific data samples, but rather learn the underlying distribution of the provided dataset.

We outlined an approach where the \ac{cnn} of a low-complexity MIMO channel estimator is trained at the \ac{bs} and just the network parameters are offloaded to the \ac{mt}. This procedure immediately resolves the problems of data generation, storage, and (re-)training of the \ac{cnn} at the \ac{mt} which is closely associated with power-efficiency constraints. Although the signaling strategy is presented for the recently proposed MIMO \ac{cnn} estimator, it may also be useful for different data-aided approaches in \ac{dl} \ac{ce}. We presented numerical results which validate the reasonability of the approach and further show that the learning-based approach outperforms state-of-the-art approaches due to the possibility of incorporating knowledge of the scenario that is available from the provided dataset.

\bibliographystyle{IEEEtran}
\bibliography{IEEEabrv,bibliography}

\end{document}